\documentclass[fleqn,10pt]{olplainarticle}
\newcommand{\ignore}[1]{}
\usepackage{hyperref}
\usepackage{wrapfig}

\title{The BaseJump Manycore\\Accelerator Network}

\author{Shaolin Xie}
\author{Michael Bedford Taylor}
\affil{Bespoke Silicon Group\\University of Washington\\ \url{http://bjump.org/manycore}}


\begin{abstract}
The BaseJump Manycore Accelerator-Network is an open source mesh-based On-Chip-Network which is designed leveraging the Bespoke Silicon Group's 20+ years of experience in designing manycore architectures. It has been used in the 16nm 511-core RISC-V compatible Celerity chip~\cite{Davidson_Celerity_IEEE_Micro_2018}, forming the basis of both a 1 GHz 496-core RISC-V  manycore and a 10-core always-on low voltage complex. It was also used in the 180nm BSG Ten chip, which featured ten cores and a mesh that extends over off-chip links to an FPGA. 
To facilitate use by the open source community of the  BaseJump Manycore network, we explain the ideas, protocols, interfaces and potential uses of the mesh network. We also show an example with source code that demonstrates how to integrate user designs into the mesh network.
\end{abstract}

\begin{document}

\flushbottom
\maketitle
\thispagestyle{empty}

\section*{Changelog}

\begin{tabular}{|l|r|} \hline
1.1 & Network Analysis Added. \\
1.0 & First Release Added. \\ \hline
\end{tabular}

\section*{Introduction}
In this document, we describe how to integrate an accelerator into the BaseJump Manycore Network, with the assumption that reader has already read the Celerity paper~\cite{Davidson_Celerity_IEEE_Micro_2018}.  This accelerator can be integrated into the middle of the manycore mesh network, or onto the south side of the network, where I/O devices are often placed. Currently, the topology that is supported by the system is a mesh topology, although it could be generalized to other topologies.

\begin{wrapfigure}{r}{0.45\textwidth}
\vspace{-3mm}
\centerline{\includegraphics[width=\linewidth]{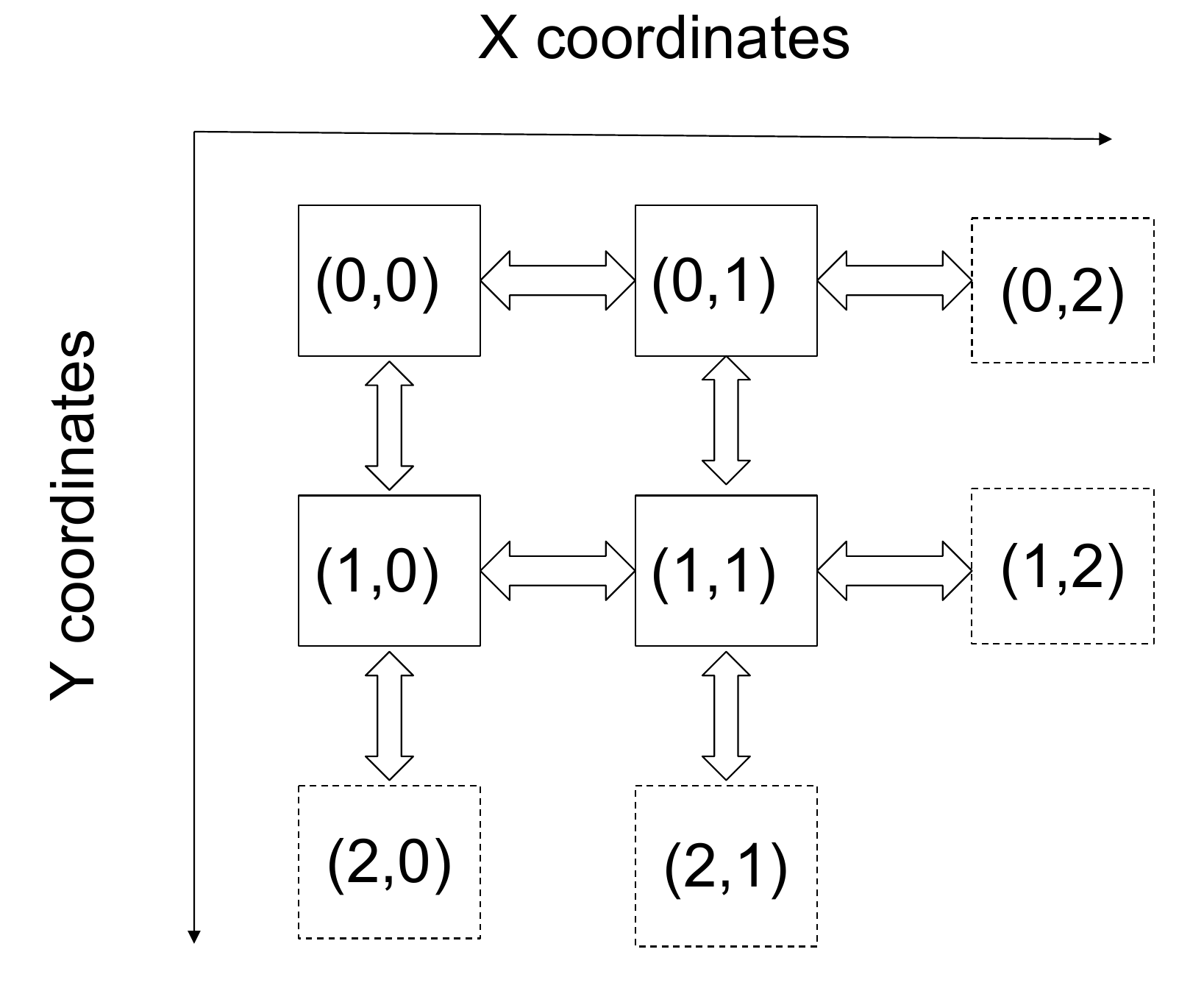}}
\caption{Global coordinates of nodes.}
\label{bsg-mesh-cord}
\vspace{-3mm}
\end{wrapfigure}

The BaseJump Manycore Accelerator-Network supports a PGAS-like (partitioned global address space) shared memory model, where a single global memory space is shared by all nodes on the network. The global memory space is addressed by the combination of a node’s XY coordinates and a local address inside that node:

\begin{verbatim}
<X cord, Y cord, local address>
\end{verbatim}

Figure~\ref{bsg-mesh-cord} shows the direction of the  X/Y coordinates. How to decode the \emph{local address} depends on the each node; it can be mapped to a memory, or simply some registers, or some other address-disambiguated functionality. The conception of the \emph{tile} or \emph{node} is very general, it can be a RISC-V core, DSP,  eFPGA, Special Accelerator, or even just a memory buffer, as shown in Figure~\ref{mesh-arch} (b).
Each tile will be allocated a local address space with the same size that is determined by the address width parameter. We call this local address space a \emph{memory region}.

This memory model allows nodes to perform load, store, and compare operations to the local address spaces of remote nodes. In prior implementations, the architecture only allowed remote stores because these are the highest performance, lightest weight operation, enabling one word per cycle throughput with minimal non-blocking hardware. With only remote stores, the architecture is very good at random scatter operations, at producer-consumer computation, and at performing multi-node barriers efficiently. The load on broken reservation instruction allows a tile to stall while waiting for an incoming store to a particular address, and is an enabler for implementing the token queue synchronization method efficiently. Later, to support random gather operations, we added the capability to perform remote loads. To better support mutual exclusion (mutex), we also added remote compare and swap operations. 

Remote loads and mutex operations require round trip communication across the mesh, and can incur long latencies, depending on the number of hops. Accelerators can be designed to tolerate this latency; for example, supporting multi-threading (but at a large area cost for multiple register file sets, and contention in the local memories), or supporting multi-word transfers (at the cost of programming complexity.)  

To the extent that such mechanisms do not exist, or are insufficient to cover the latency, two sub-modes of operations are preferred. Either the code should attempt to achieve locality of remote accesses, localizing remote memory usage and reducing round trip latency; or it can perform the remote blocking operations in parallel across many tiles. In the later case, the bisection bandwidth of the network sets an upperbound on the execution resource efficiency of the core. 

For example if every core sent a message across the median of the array, with 16 links crossing the bisection, only 32 remote operations can be sustained per cycle, corresponding to one operation per 16 cycles on a core. In such cases, with a 48-cycle average round trip latency in the mesh, the relative underutilization is only 3X.

In the following sections, we overview the operation of the network that supports this memory model and how an accelerator interfaces to this network.

Generally, speaking if your accelerator does loads and stores to remote tiles and immediately processes incoming remote loads and stores to its local memory, it will be effortless to integrate using a module called {\em the standard endpoint} and relatively little understanding of the network is required.  

If your accelerator is trying to do more streaming behaviors, where it uses the packet as an streaming remote procedure call (RPC) rather than as a simple memory read/write requires, and its computation model is that it has a stream of incoming requests, and wants to take a request, process it and send it out to another accelerator on the request network \emph{before dequeing the next request from the incoming network}, then you will need to understand the network extremely well and design your accelerator to meet the requirements of the network. Examples of these kinds of accelerators include a DRAM controller that sits on the edge of the chip, or a signal processing or software radio filter.

\section*{Mesh Network Overview}
\begin{figure}[tb]
\centerline{\includegraphics[width=\linewidth]{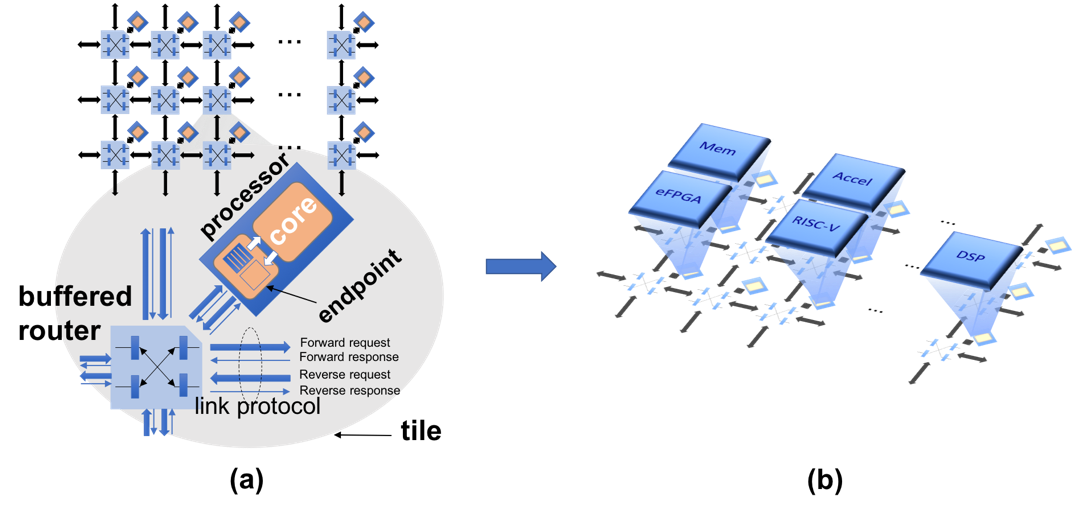}}
\caption{Mesh Network : Accelerators or processors are attached to the 4-direction routers. The routers use \textit{link protocol} to communicate.}
\label{mesh-arch}
\end{figure}

Figure~\ref{mesh-arch} shows the mesh architecture. Each tile contains a router and an accelerator (labeled processor in the figure), and each accelerator contains an endpoint and a core. Messages on the network are a single wide word that includes header information and payload. 

{\bf \noindent Network Composition and Usage Restrictions.}
The \textit{link protocol} used in the mesh network contains an input link and an output link for each direction. Each link uses independent \textit{forward} and \textit{reverse paths}. Request packets travel from source tile to destination tile on the forward path; and reply packets from destination to source occurs on the reverse path. All messages on the forward path have replies; for example a store request packet results in a credit packet on the reverse network that indicates that the store has commit. 

Accelerators \emph{should} be designed so that they can absorb all incoming messages presented by the request network router immediately upon arrival and are \emph{forbidden} from making the absorption of incoming messages dependent on sending out new messages on the request network. This prevents messages from clogging up the routers, leading to congestion in the network, greatly reducing performance. There are two ways to do this. The first is to design an interface that can service incoming packets at line rate of 1 word per cycle; for example an interface that accepts only the standard load or store operations to a local SRAM, which can be easily serviced \footnote{A clever rule is to prefer give priority to local memory accesses over arriving remote memory accesses until the incoming FIFO fills up; this opportunistically waits for spare slots and minimizes the disruption of the remote memory accesses up until the point it starts backing up into the network.}  and will not result in cache misses that would reduce this line rate. The second is to provide a large enough input FIFO to buffer accumulating incoming messages from all possible parties that could result from non line-rate servicing. For unrestricted communication patterns where any node can send to any other node, this results in $N^2$ words of FIFO buffering being inserted chip-wide per outstanding transaction, which may be an unnecessarily over-provisioned scenario.

Violating the rule of absorbing incoming packets immediately on the forward path is not strictly required to avoid deadlock, since there is a separate path for replies. However, it can lead to terrible chip-level performance, because the packets will back up into the network and cause congestion quickly. To address the issue of avoiding back up into the network, there are three preferred options, and one non-preferred option:

Option 1: Use the \emph{token queue} primitive to virtualize channels between nodes using a load/store shared memory interface. This allows for multiple input or output channels across nodes, with flow control.  Downside: complexity.

Option 2: In cases like a pipeline of filters, where one node is only sending data to a second node, and there are no other messages going to the second node, the first node can have an outstanding message counter that causes it to stall when the number of outstanding messages is equal to the size of the second node’s input FIFO. The size of the FIFO can be set to optimize performance. Downside: works only for one-to-one communication pipes.

Option 3: If it can be proven that the backup is extremely unlikely, the performance degradation can be shown to be acceptable vis-a-vis the cost of enlarged FIFOs. Downside: complicated proof, possibility of unexpected worst-case performance.

Option 4 (highly discouraged): allow messages to backup into the network, blocking other traffic, and statically ensure that no other tiles are using those network links.

Allowing messages on the forward path to block because of reverse network blockage is permitted because we want to make the amount of buffer space in the forward path small and finite for the common case of remote shared memory operations, and because if routing is done correctly, the chances of delays are slim.
 
All messages on the reverse path \emph{must} be absorbed immediately upon arrival at its final destination to ensure that there are no cycles in the collective networks. This is an easy requirement to fulfill, because any core that initiates a message should be able to provision enough space to receive the responses. However, it does mean that the receiver cannot, for example, condition the receipt of the response on being able to send a message on the request network. By design, we omit backwards flow control on this interface; it cannot block at all, for example, waiting on an outgoing message. Because of this, we can prove the network can never deadlock because messages will always be sunk on the reverse path, causing messages to move forward on the forward path. (Refer to~\cite{taylor2007tiled}  Section 2.8 for a more detailed proof. The independent response network in the mesh is a 'sink' network in which each node will never block and can always absorb the response. This independent 'sink' network make the mesh network acyclic, which means there will never be any interdependence paths in the network, making it 'deadlock-free'.)

{\bf \noindent Deadlock-free Communication:} 

{\bf \noindent Endpoint Interface.} 
Adherence to the protocol rules of the network is not always obvious to designers who are adding new kinds of nodes to the network. For these reason, we provide two hardware blocks: the \textit{standard endpoint} and the \textit{barebones endpoint}. The standard endpoint is more full-featured and supports the deadlock-free protocol, credit flow control and other special request like start/stopping the core, atomic swap operation, remote loads and stores and store barriers. The rest of this document details the usage of the standard endpoint. The barebones endpoint is the barest bones interface; it buffers the incoming forward path with a small input FIFO of configurable side, and decodes the network links into local handshakes used by the local core. It requires additional logic to make it compliant with the mesh network requirements but is useful for implementing very lightweight hardware accelerators in the network.

The benefit of this separation is providing a standard plug-and-play interface for the core, in which the core can see the network as a general master/slave module and do not have to concern about the flow control/buffer problems that related to the network. 

{\bf \noindent Packet Based Inter-tile Synchronization.}
Hard-wired synchronization networks across tiled architectures can be easy to implement and ultra low latency but struggles to meet the demands of software; namely to be able to support arbitrary numbers of nested synchronizations comprising arbitrary subsets of non-contiguous tiles.  Instead, we use packet-based inter-tile synchronization and also a built-in atomic compare-and-swap function. 
Other high level primitives like mutex, barrier, and spin-lock can layer on top of the built-in atomic compare-and-swap function, and these primitives can distributed on any tile at runtime, giving more flexibility on task decomposition and mapping. 
Furthermore, the high level primitives can either be implemented with software (if it is a processor) or hardware (if it is an accelerator), providing more flexibility on design trade-offs.

\section*{Operational Details}
\subsection*{Transaction Delay}
\begin{figure}[tb]
\centerline{\includegraphics[width=\linewidth]{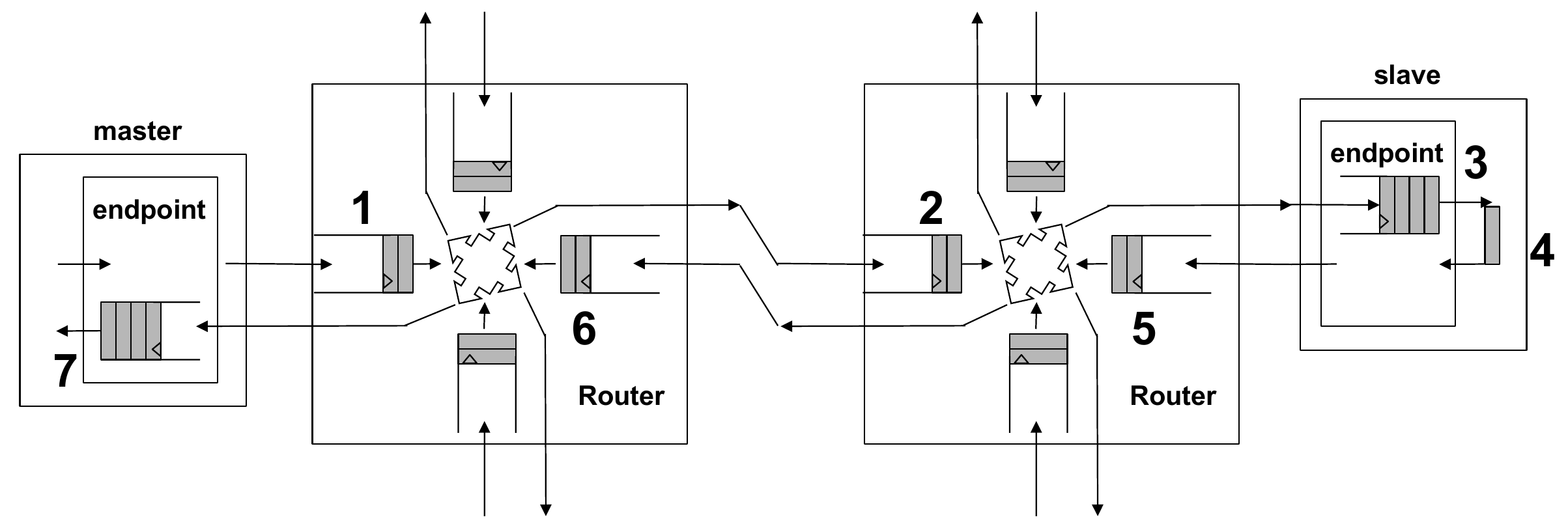}}
\caption{Example round trip delay of the mesh network.  The number indicates the transaction path and also the incremental delay. If there is network congestion, the delay at router will increase due to both round-robin arbitration and head-of-line blocking from a neighbor router.  The router uses round-robin arbitration, so the router arbitration delay varies between 1 and 5 (5 is the number of directions).}
\label{network_delay}
\end{figure}

Figure~\ref{network_delay} shows the transaction path and the related delay in an unloaded network.  Each time a packet crosses a FIFO, a 1 cycle delay is added. Routers have input FIFOs and no output FIFOs. Endpoints tend to have input FIFOs, which are configured based on the traffic it is expected to receive.

In the \textit{mesh\_master\_example.v} (See Subsection entitled \emph{Attaching Master Modules}), there is a counter starting at the cycle that issue the first read request, and a monitor that print the counter value and the returned value. As we can see from the output, the delay of the first response is exactly as 7 clock cycle.
\begin{verbatim}
cycle 7, returned=00000000, expected=000
cycle 8, returned=00000001, expected=001
cycle 9, returned=00000002, expected=002
\end{verbatim}

\subsection*{Store Credit Counting}

\begin{figure}[tb]
\centerline{\includegraphics[width=\linewidth]{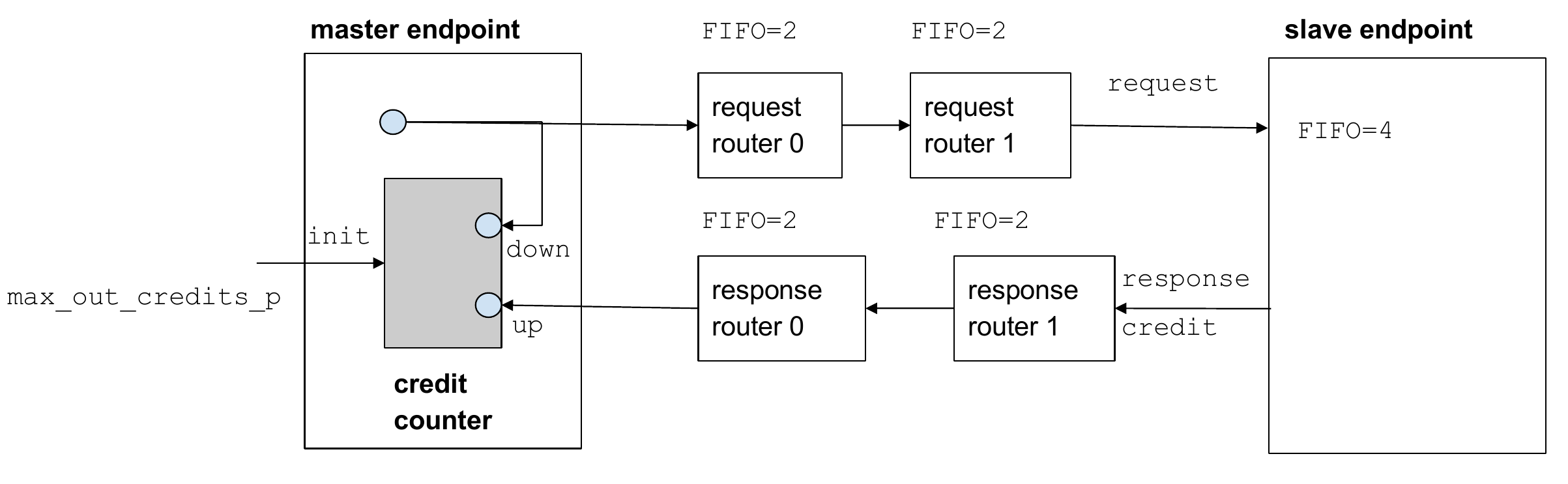}}
\caption{Credit based flow control.  In the BaseJump mesh network, each transaction will be acknowledged with a credit via the independent 'reverse' response network. The acknowledgement does not just mean that the packet has arrived, but that it has been irreversibly committed and no subsequently received packet from anywhere in the machine can jump in front. This requires certain implementation guarantees at the endpoint. FIFO sizes are indicated in the diagram.}
\label{flow_control}
\end{figure}

The BaseJump mesh network uses a credit mechanism, shown in Figure~\ref{flow_control}, primarily to be able to implement store barriers that indicate that all stores have committed at their final destination; but this mechanism is also used to bound how many outstanding packets a node may have. 

In the standard endpoint hardware module, a credit counter will track the number of the outstanding transactions that has not been acknowledged yet. The counter is initialized with the parameter \texttt{max\_out\_credits\_p}. This is used for implementing memory barriers.

Typically, we want to set the number of outstanding credits to be greater than the uncongested bandwidth delay product of the longest roundtrip path. For example, if we can issue 1 word per cycle stores to a memory on the opposite corner of the chip that has a 128-cycle latency from sending the store to receiving the store acknowledgement back, then we want to set it to at least 1 word/cycle * 128 cycles = 128 credits.

Generally there should be little advantage to setting it much higher than this amount, except to tolerate post-fabrication changes in delays to external devices (e.g. DRAM), and the disadvantage is that during congestion, the network may get further clogged with unnecessary packets.

In streaming communication cases where the communication patterns between nodes are one-to-one; max outstanding credits can be set to the FIFO capacity of the destination node to eliminate backup into the network. (As mentioned before, for non one-to-one communication, even with bounded credits, all of the cores could potentially transmit to the same end node, avoiding congestion would require that the destination buffer to be extremely large to be able accommodate all possible packets from all possible senders; in this case the token queue mechanism should be used.)  

\ignore{
{\bf \noindent Maximum outstanding transactions:} The maximum number of outstanding transactions is determined by the collective FIFO depth of the network paths in an uncongested scenario. For example, in Figure~\ref{flow_control} the FIFO depth is 2 for router, and 4 for endpoint. So the total FIFO depth is (2 + 2+ 4 + 2 + 2 =12). Suppose there is a congestion at ‘response router 0’,  the master endpoint can issue 12 transactions before being blocked at ‘request router 0’. However, this usage is highly discouraged: in-network buffer space should not be relied upon for buffering; instead the endpoint FIFOs should be sized appropriately to absorb all incoming packets and eliminate unnecessary congestion in the network.

Though congestion will not cause deadlock, it will impact the effective bandwidth.  Setting \texttt{max\_out\_credits\_p} to a value between minimum and maximum outstanding transactions, and stop sending request when run out of credit, will allow some congestion but prevent traffic jams.
}
\subsection*{Transaction ordering}
\begin{figure}[tb]
\centerline{\includegraphics[width=0.5\linewidth]{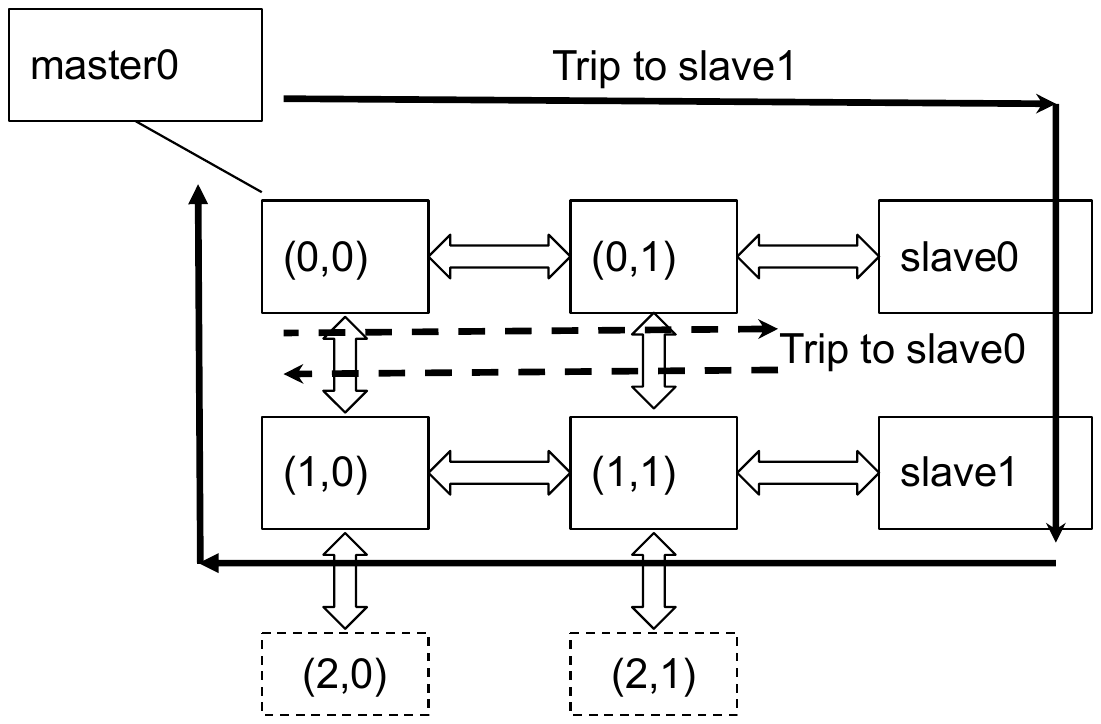}}
\caption{Transaction ordering example: If master 0 issues a load request to slave1 and then a load request to slave 0, the response from slave 0 will reach master 0 before that of slave1, causing out-of-order transactions.}
\label{trans_order}
\end{figure}

The network is an ordered network, which means that all packets sent by a source node to the same destination node must arrive in order. Each destination node must also ensure point-to-point ordering on load or store request; i.e. load or store requests received at a destination node from a particular source node should be committed in sequential order. There is no guarantee of ordering between different nodes. Figure~\ref{trans_order} shows a case when the transaction may be completed out-of-order.

{\bf \noindent Transaction Fence:} if the master wants to implement transaction fence which only issue transactions after previous transactions are all completed, it need to wait until the credit counter value back to ‘max\_out\_credits\_p’. The bsg\_manycore\_endpoint\_standard module provides a dedicated output ‘out\_credits\_o’ indicates remaining credits. 

\subsection*{Routing Constraints}
BaseJump mesh network uses XY dimension ordered routing, which means the transaction will first travel along one dimension (e.g. the X dimension), and then travel along the other dimension (e.g. Y dimension). Dimensioned-ordered routing is a simple way to prevent deadlock inside the network, but does not prevent \emph{endpoint deadlock}; instead that must be realized by placing restrictions on how the nodes use the network.

\begin{figure}[bt]
\centerline{\includegraphics[width=\linewidth]{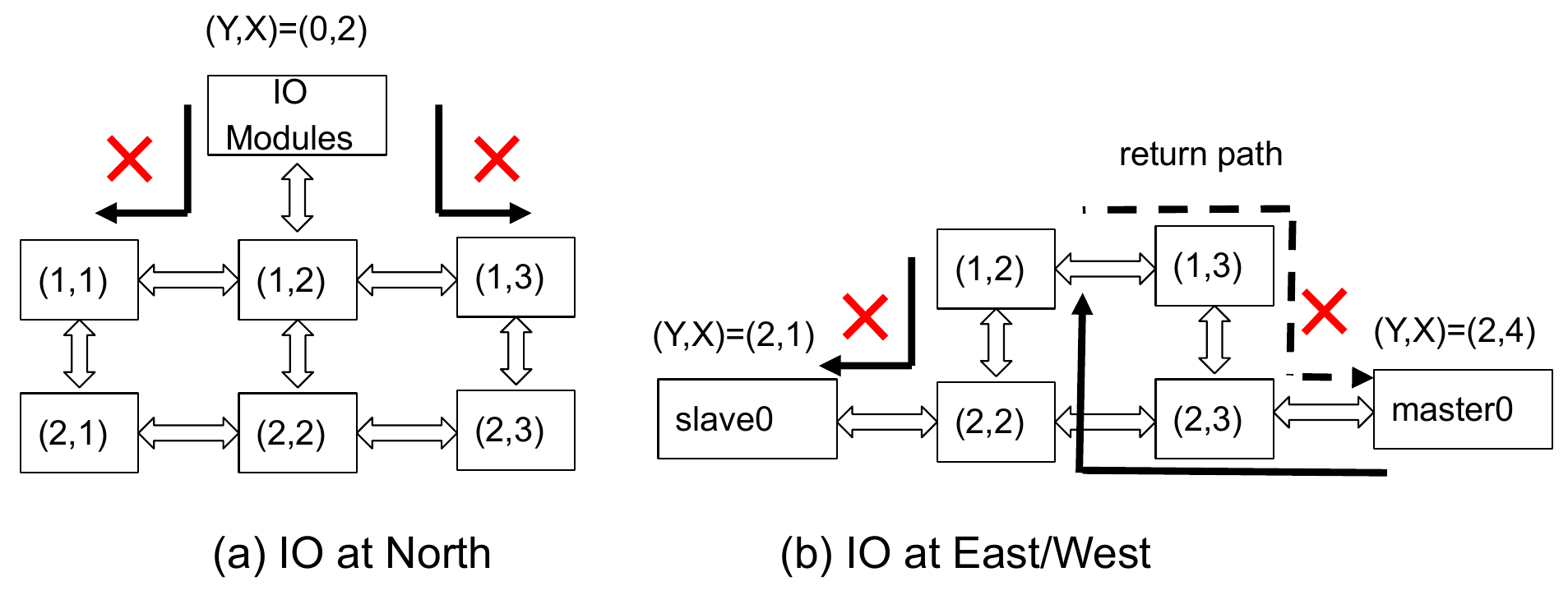}}
\caption{IO can only be connected to south of the mesh network.}
\label{routing_const}
\end{figure}
The router has 5 input directions and 5 output directions (P=Processor, W=West, E=East, N=North, S=South). If we support all-to-all routing, there will be a large 5x5 crossbar. To reduce the resource usage, we leverage the fact that dimension-ordered routing typically does not have routes that go from North/South to East or West. However, to support I/O on the edges, we do allow routing from south to east or west. 

The following routing is not allowed:
\begin{itemize}
\item \textbf{N $\rightarrow$ W}
\item \textbf{N $\rightarrow$ E}
\end{itemize}

With these prerequisites, the following constraints apply for the BaseJump mesh network
\begin{itemize}
\item IO module can only be attached on the south boundary or as a accelerator node
\end{itemize}

\subsection*{Virtual Mesh, Sub Mesh, and Sub Node}

The biggest advantage of mesh network is that it can be directly mapped to physical layout, with constant wire lengths, which guarantees the scalability and feasibility.
\begin{figure}[bt]
\centerline{\includegraphics[width=\linewidth]{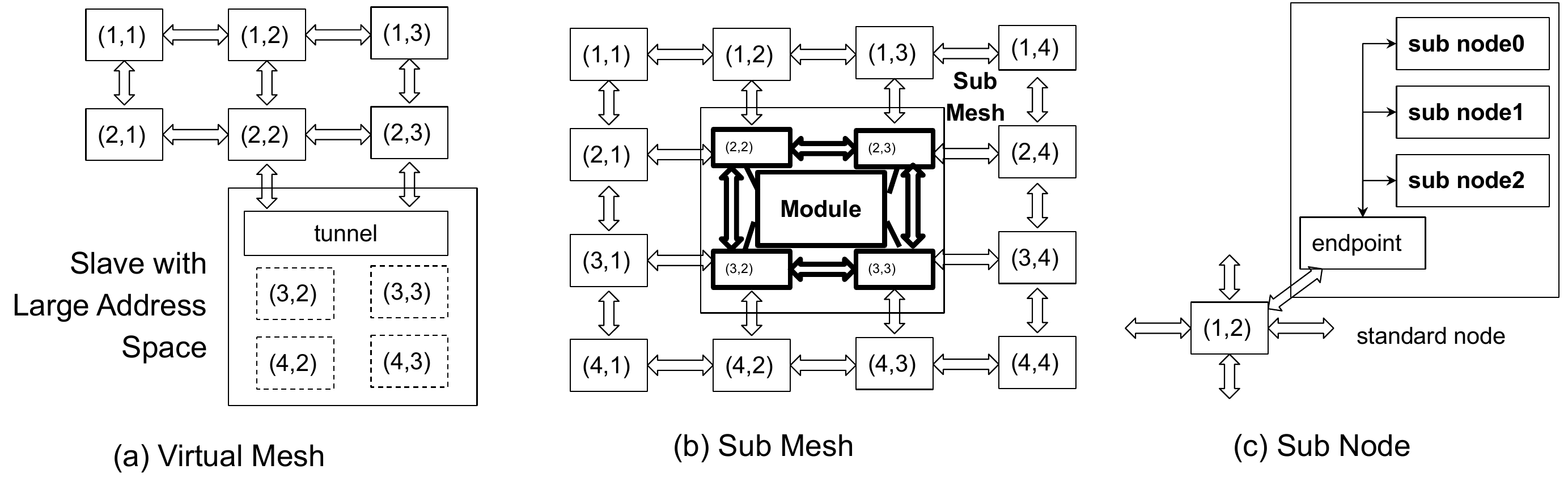}}
\caption{Virtual Mesh, Sub Mesh and Sub Node}
\label{hyper_mesh}
\end{figure}

However, in a hetero system with various accelerators, the physical size of nodes are usually not the same, causing irregular  physical layout and thus large amount effort in physical design and tuning.

To maintain the regularity of mesh network and scalability, we propose the concept of virtual node, super node and sub node, which address the large IO, large accelerator and small accelerator problem respectively. 

{\bf \noindent Attaching Large IO as Virtual Mesh:} For module with large address space, for example, DRAM controller, we can allocate multiple nodes for it. The cheapest way to do this is to have additional Y coordinates available to multiply up the amount of space at the periphery. X coordinates can also be used, but this requires hardware to merge the streams.  In cases where we are virtualizing over a single link  \textit{bsg\_channel\_tunnel} or a round robin module can be used to merge the links. 

There is no actually mesh for  memory regions inside the DRAM controller, but the traffic will still be routed the mode correctly. We call these conceptual node ‘virtual mesh'. Figure~\ref{hyper_mesh} (a) shows conceptual fabric to integrating DRAM controller with virtual mesh.

{\bf \noindent Attaching Large module as Sub Mesh:} For module with large physical size, e.g, a large matrix multiply accelerator, we attach it into the network as sub mesh. The sub mesh will be allocated with multiple memory regions based on its physical size. So even the module has only few memory spaces, it might be allocated with multiple memory regions if its size is large than a 'standard' node.

To maintain the XY dimensional  routing algorithm, the sub-mesh most likely will have some kind of routers implemented, as shown in Figure~\ref{hyper_mesh} (b). 

{\bf \noindent Attaching tiny module as Sub Node:} For module with small physical size, e.g, a floating point unit, we take it as sub node which will be encapsulated into standard nodes with other sub node. The sub node will be allocated with a sub memory region, and an extra address mux will be used to route traffic between the sub node and the endpoint.

\section*{Attaching Basic Master/Slave Modules}
\label{sec:attaching}
\begin{figure}[tb]
\centerline{\includegraphics[width=0.8\linewidth]{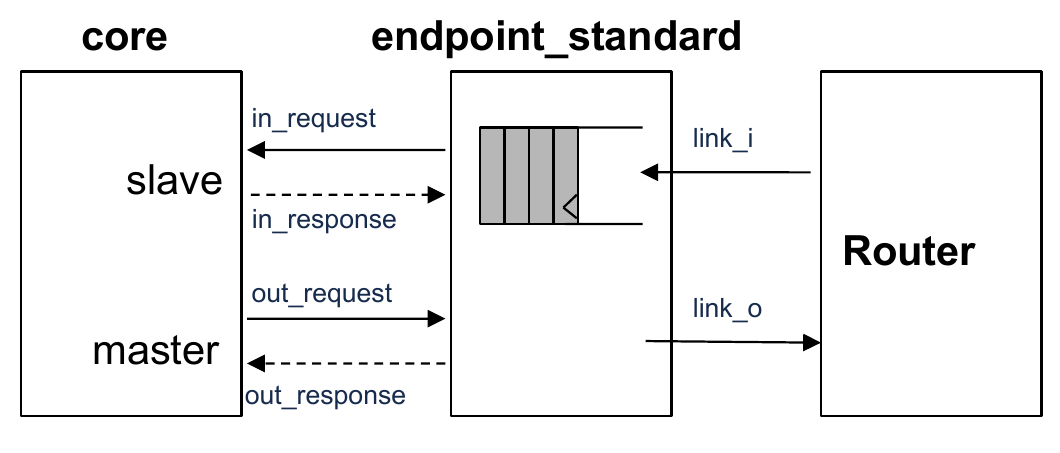}}
\caption{endpoint\_standard interface: The \textit{link\_i} and \textit{link\_o} are signals interface with router only, the core module do not have to care about that.}
\label{endpoint}
\end{figure}

In this section we will show how to integrate basic modules into the mesh network. Any module attached to the mesh network should instantiate an \textit{endpoint\_standard} unit, which provides general master/slave interface to the core. The core can support either 'slave' or 'master' interface or both. If the core does not support master or slave interface, it can tie up the corresponding output signals and ignore the input signals. Figure~\ref{endpoint} shows the basic interface of \textit{endpoint\_standard}.

{\bf \noindent Handshakes:} Three types of latency-insensitive handshake used in the interface~\cite{taylor_basejump}.
\begin{itemize}
\item {\bf valid and ready:}
The sender pull up ‘valid’ signal whenever data is available, and the receiver pull up ‘ready’ signal whenever it can receive data. 

When both ‘valid’ and ‘ready’ are pull up at the same cycle, the transaction is done. Both valid and ready should be asserted near the beginning of the cycle.

\item {\bf valid and yumi:}
The sender asserts ‘valid’ signal if data is available, and the receiver should only pull up the ‘yumi’ signal when it  see the ‘valid’ signal and it can receive data. 

When ‘yumi’ is pulled up, the transaction is done. Valid should be asserted towards the beginning of the cycle, and yumi can be asserted anytime until the end of the cycle.

\item {\bf Valid:}
Here, valid is asserted towards the beginning of the cycle, and the receiver must always receive the data. 
\end{itemize}

\subsection*{Attaching Slave Modules}
A slave attached to the network supports general load/store operations that come in from the network.

{\bf \noindent in\_request:} load/store request from the network.
\begin{itemize}
\item \textit{fields:} local address, store data, store mask, and write enable.
\item \textit{handshake:} valid and yumi. The slave yumi’s the request from network.
\end{itemize}

{\bf \noindent in\_response:} load result send to the network.
\begin{itemize}
\item \textit{fields:} loaded data
\item \textit{handshake:} valid. To avoid complexity inside the slave module,the endpoint only forwards ‘in\_request’ when reverse channel is available, otherwise, the ‘in\_request’ will be masked. So endpoint is always capable of absorbing the response.
\end{itemize}

{\bf \noindent Rules:} The in\_response should be returned at least one cycle after the corresponding in\_request. 

{\bf \noindent Example:} \url{bsg_manycore/testbenches/mesh_example/mesh_slave_example.v} shows how to connect a general memory to the mesh network. Some notes:
\begin{itemize}
\item {\it \noindent yumi} signal: 
\begin{verbatim}
//we can always handle the request
assign  in_yumi_li  =       in_v_lo   ;     
\end{verbatim}
In this particular example, the memory can always handle the read/write request so we consume the request when it is available.
For other more complex slave like memory controllers or memory+cache controllers, the yumi signal may only be pulled up when slave is ‘idle’ and the request is available 
\footnote{Ideally the memory controller will have enough input FIFO space to allow messages to be pulled from the network in most cases, eliminating congestion. Since the possible number of outstanding requests to a memory controller grows with the number of tiles, this could be as much as N packets where N is the number of tiles. For 512 cores, this could be 5 kbytes of memory per memory controller. Not terrible, but the system would also likely work well if it were $2\frac{N~cores}{M}$ under the assumption of random load balancing across M banked memory or cache controllers.
}.

\item {\it \noindent returing\_v\_i} signal: 
\begin{verbatim}
//the returning data is only available when it is a read request
    Always_ff @(posedge clk_i)
        if( reset_i ) returning_v_r <= 1'b0;
        else          returning_v_r <= (in_yumi_li & ~in_we_lo);
\end{verbatim}
In this particular example, the returning\_v\_i is a delayed read request. For write request, there is no returning data.
For other complexed slave, for example, memory controller, the returning\_v\_i signal may be asserted with multiple cycles of delay.
\end{itemize}

\subsection*{Attaching Master Modules}
\label{master-modules}
A master attached to the network only has to support sending out remote load/store/atomic operations.

{\bf \noindent out\_request:} load/store request to the network.
\begin{itemize}
\item \textit{fields:} X Y cord, local address, store data, store mask, and write enable.
\item \textit{handshake:} valid and ready.
\end{itemize}

{\bf \noindent out\_response:} load result send to the network.
\begin{itemize}
\item \textit{fields:} loaded data
\item \textit{handshake:} valid. To avoid complexity inside the slave module,the endpoint only forwards ‘in\_request’ when reverse channel is available, otherwise, the ‘in\_request’ will be masked. So endpoint is always capable of absorbing the response.
\end{itemize}

{\bf \noindent Rules:} The master must receive the returned data immediately when it appears. 

{\bf \noindent Example:} \url{bsg_manycore/testbenches/mesh_example/mesh_master_example.v} shows how to issue requests to the mesh network and receive the response.  In this example, the master will first write a sequence of data into a memory region and then read the data back. The address, written data and how many read/write commands are controlled by state machine. Refer to state machine code section for more details. 
For sending request, we only have to assign these two structs:
\begin{itemize}
\item {\it \noindent out\_v\_li} signal: 
\begin{verbatim}
assign out_v_li = (stat_r == eWriting) || (stat_r == eReading)   ;
\end{verbatim}
In this particular example, we only issue the request in writing and reading state.

\item {\it \noindent out\_packet\_li} signal: 
\begin{verbatim}
assign out_packet_li = '{
                   addr           :       addr_r
                  ,op             :       eOp_n
                  ,op_ex          :       {(data_width_p>>3){1'b1}}
                  ,data           :       data_r
                  ,src_y_cord     :       my_y_i
                  ,src_x_cord     :       my_x_i
                  ,y_cord         :       dest_y_i
                  ,x_cord         :       dest_x_i
                  };

\end{verbatim}
The outgoing request is  encapsulated in a SystemVerilog struct. Please refer to \textit{Appendix A: Endpoint interface} for detailed explanation for each field. Generally we use struct syntax to assign value as shawn in the above code snippet.
\end{itemize}

For returned data, we only have to monitor these two output from endpoint\_standard:
\begin{verbatim}
   ,.returned_data_r_o(  returned_data_lo      )
   ,.returned_v_r_o   (  returned_v_lo         )
\end{verbatim}

Loads and stores to the same memory region (the same x\_cord, y\_cord) are ordered, but that is no guarantee order for datum from different memory region unless the system waits for all prior transactions to complete. Please refer to \textit{Transaction ordering} section for more discussion about the transaction ordering.

\subsection*{Toplevel Connection}
At top level, we use \texttt{bsg\_manycore\_mesh\_node} to connect the master and slave. Here is an example topology:
\begin{verbatim}
//
//  MASTER(0,0)   TIED          TIED       TIED      (Y, X)
//        \        |               \        |
//         \       |                \       |
//          |-------------|          |-------------|
//   TIED---| ROUTER(0,0) | ---------| ROUTER(0,1) | --- TIED
//          |-------------|          |-------------|
//                 |                        |
//                 |                        |
//               TIED                     SLAVE(1,1)

\end{verbatim}
Each mesh\_node have 5 directions (Attached Processor, West, East, North, South) that is hard coded in \texttt{bsg\_noc\_pkg}: 
\begin{verbatim}
typedef enum logic[2:0] {P=0, W, E, N, S} Dirs;
\end{verbatim}
Generally we use these hard coded directions to index the signal arrays of the node. For directions that have nothing attached, we need to stub that direction and also attached tie offs to the signals. 
\begin{itemize}
\item {\it \noindent stub\_p parameter:} stub\_p is a 4-bits vector corresponding to the 4 directions of the node (The core direction is always enabled). When the corresponding bits is set, the direction is stubbed and will not forwarding packets. The main purpose of the parameter is to save the resource (e.g buffers) when the direction is not connected.

\item {\it \noindent tieoff module:} for unconnected direction, we use \textit{bsg\_manycore\_link\_sif\_tieoff.v} module to tie off.
\end{itemize}

\section*{Appendix A: Endpoint interface}
Source code: \url{bsg_manycore/v/bsg_manycore_endpoint_standard.v}

{\noindent Parameters:}
\begin{itemize}
\item x\_cord\_width\_p:  	the width of the X coordination, defined by system
\item y\_cord\_width\_p: 	the width of the Y coordination, defined by system
\item fifo\_els\_p:		 the FIFO depth inside endpoint, usually 4.
\item data\_width\_p:		usually 32.
\item addr\_width\_p:		local address bit width (in WORDS), usually 20
\item max\_out\_credits\_p:	How many packet can be send out without return credit.
			Determined by the round-trip hops multiplied by FIFO 
			depth. For example, if the maximum hops is 20 and FIFO 
			depth is 4, we can set this parameter to 20x4=80.
\item warn\_out\_of\_credits\_p: print warning message if out of credit
\item freeze\_init\_p		 :  after reset, the freeze\_r\_o signal will be set to this value
\end{itemize}

\subsection*{Signal Groups}
{\bf \noindent in\_request:} incoming request from the network.
\begin{itemize}
\item in\_v\_o:  	valid signal
\item in\_yumi\_i: 	yumi signal to consume the incoming data
\item in\_data\_o:	incoming data
\item in\_mask\_o:	incoming mask, 1 bit mask for each 8-bit data.
\item in\_addr\_o:	incoming local address (in \textbf{WORDS})
\item in\_we\_o:	write enable.
\end{itemize}

{\bf \noindent in\_response:} response for the incoming request. Will send to the network.
\begin{itemize}
\item returning\_v\_i:		returning data is valid 
\item returning\_data\_i:  	returning data
\end{itemize}

{\bf \noindent out\_request:} outgoing request to the network, using ‘valid and ready’ handshake.
\begin{itemize}
\item out\_v\_i:  	valid signal
\item out\_packet\_i: 	outgoing packet
\item out\_ready\_o:	ready signal
\end{itemize}

The outgoing packet format is defined in \url{bsg_manycore/v/bsg_manycore_packet.vh}, which include following field.  Usually we use \textit{bsg\_manycore\_pkt\_encode.v} to generated the packet.
\begin{verbatim}
  typedef struct packed {                           \
      logic [(in_addr_width)-1:0]    addr;  //(in WORDS) \
      logic [1:0]                    op;             \
      logic [(in_data_width>>3)-1:0] op_ex;          \
      logic [(in_data_width)-1:0]    data;           \
      logic [(in_y_cord_width)-1:0]  src_y_cord;     \
      logic [(in_x_cord_width)-1:0]  src_x_cord;     \
      logic [(in_y_cord_width)-1:0]  y_cord;         \
      logic [(in_x_cord_width)-1:0]  x_cord;         \
   } bsg_manycore_packet_s
\end{verbatim}

The meaning of each field are defined as following:
\begin{itemize}
\item op	:  request type, supported request are defined as following:
\begin{verbatim}
`define  ePacketOp_remote_load    2'b00
`define  ePacketOp_remote_store   2'b01
`define  ePacketOp_remote_swap_aq 2'b10   // used for atomic swap only
`define  ePacketOp_remote_swap_rl 2'b11   // used for atomic swap only
\end{verbatim}
\item op\_ex	:  set to mask for store operations
\item src\_y\_cord	: Y cord of current tile
\item src\_x\_cord	: X cord of current tile
\item y\_cord		: Y cord of destination tile
\item x\_cord		: X cord of destination tile
\end{itemize}

{\bf \noindent out\_response:} NO HANDSHAKE, THE CORE MUST ACCEPT THE DATA
\begin{itemize}
\item  returned\_v\_r\_o:  	valid signal
\item  returned\_data\_r\_o: 	returned data
\end{itemize}

{\bf \noindent control signals:}
\begin{itemize}
\item out\_credits\_o:  How many credits available.
\begin{itemize}
\item Congestion Control:The core should avoid to send request if out of credit.
\item Fence control: If the core need to wait all outstanding request finish, it can just wait the credits equal to \textit{max\_out\_credits\_p}.
\end{itemize}

\item my\_x\_i: 		X cord of current tile
\item my\_y\_i: 		Y cord of current tile
\item freeze\_r\_o:		A freeze packed is received
\begin{itemize}
\item 0 :		The tile should be frozen 	(stopped)
\item 1 :		The tile should be unfrozen	(started  )
\end{itemize}
\item reverse\_arb\_pr\_o:	Reverse the priority of the arbiter inside RISC-V. Can be
\end{itemize}

\subsection*{Special Local Address Map}
\begin{wrapfigure}{r}{0.3\textwidth}
\centerline{\includegraphics[width=0.9\linewidth]{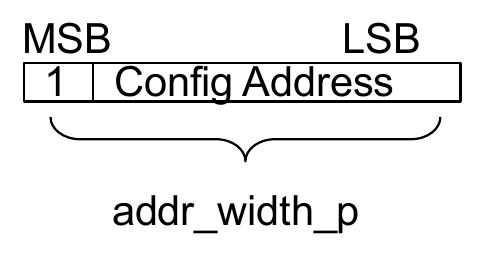}}
\caption{Configuration Address Space, where MSB is tied to 1.}
\label{addr_space}
\end{wrapfigure}

Right now, we defined an special 'configuration' local address space and the decoding is hard coded inside endpoint, as shown in Figure~\ref{addr_space}. 

{\noindent Configuration Registers:}
\begin{itemize}
\item 0x0: 		Freeze Register
\begin{itemize}
\item Value=0,  unfreeze
\item Value=1,  freeze
\end{itemize}
\item 0x4: 		Priority Arbiter Register
Each write operation will toggle the value.
\end{itemize}

\section*{Network Analysis}

The default mesh network implementation is light-weight and designed to minimize area and energy consumption, enabling maximum density for relatively small tiles. The buffer sizes are minimal, the router has limited path diversity, and there are no attempts to manage congestion inside the network. For designs with larger nodes, or with particularly adversarial traffic patterns, it may make sense to add other features (e.g. larger buffers, FIFOs used when turning in the dimension-ordered routers to reduce contention, express routes, virtual, physical, or logical channels, etc.), after doing a detailed trade-off analysis.

Here we analyze the default configuration, under uniform random traffic, where each node has equal probability of sending to each other node.

Suppose we have a $k$ by $k$ mesh of tiles. The limiting minimum bisection of the mesh under this workload is $k$ words per cycle (one per network link), going across the center of the array of tiles.  Under uniform traffic, half of the nodes ($k/2$) have a $1/2$ probability of sending to the other side of the chip, going over the mesh bisection. Therefore, for each round of packets sent, we would have $k/2*k/2=k^2/4$ packets that need to cross the bisection. Since we have $k$ network links on the bisection, this traffic would occupy each link for $\frac{k^2}{4} / k = \frac{k}{4}$ cycles. So for an optimal $16-by-16$ network, the nodes would be able to inject a new message every $16/4 = 4$ cycles.

Below are some example figures for the performance of a dimension-ordered routed (DOR) $8$-by-$8$ 2D mesh and are compared to other routing algorithms like Valiant (VAL), ROMM, and MAD (Minimal Adaptive Routing). These are from Dally and Towle's highly recommended textbook, ``Principles and Practices of Interconnection Networks.'' They show three types of traffic, uniform random, transpose, and nearest neighbor. To be clear, these graphs are not for our particular network, but are qualitatively representative.

These graphs have a standard format that is used by the interconnection network research community. The graph measures the average packet delay versus the offered traffic, which is the rate at which new packets are launched by the nodes. Offered traffic is normalized to the best case for that traffic pattern. So for example, the uniform case is normalized to the 0.25 messages per cycle limit set by the bisection bandwidth of an $8$-by-$8$ mesh. Packet delay is measured in a subtle fashion. The model assumes an infinite queue at the sender, so that a new word can be enqueued at a fixed rate even if the network is backed up. As you can see, at some point when the offered traffic approaches the upper bound on traffic, depending on the communication pattern, then the delay of delivery of the packet spikes up because the network is congested. Transpose traffic is challenging for DOR, while nearest neighbor is very efficient.

\begin{figure}[tb]
\centerline{\includegraphics[width=\linewidth]{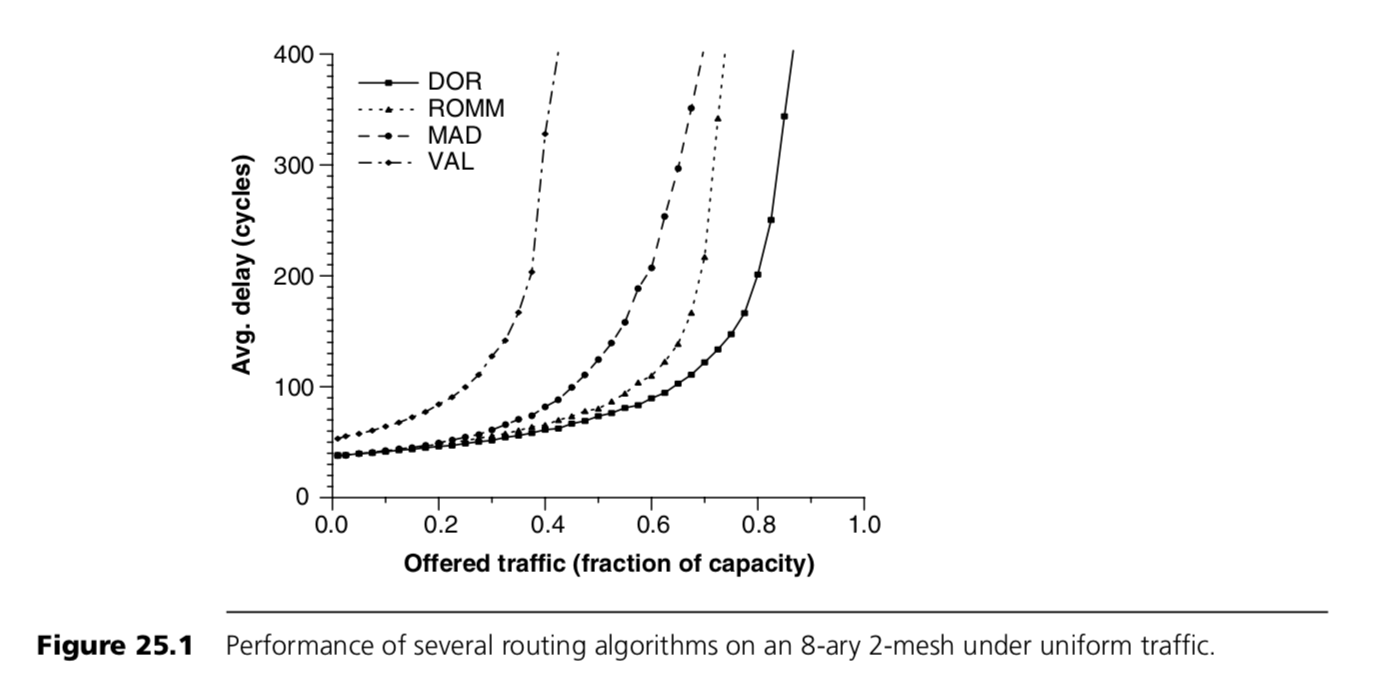}}
\end{figure}

\begin{figure}[tb]
\centerline{\includegraphics[width=\linewidth]{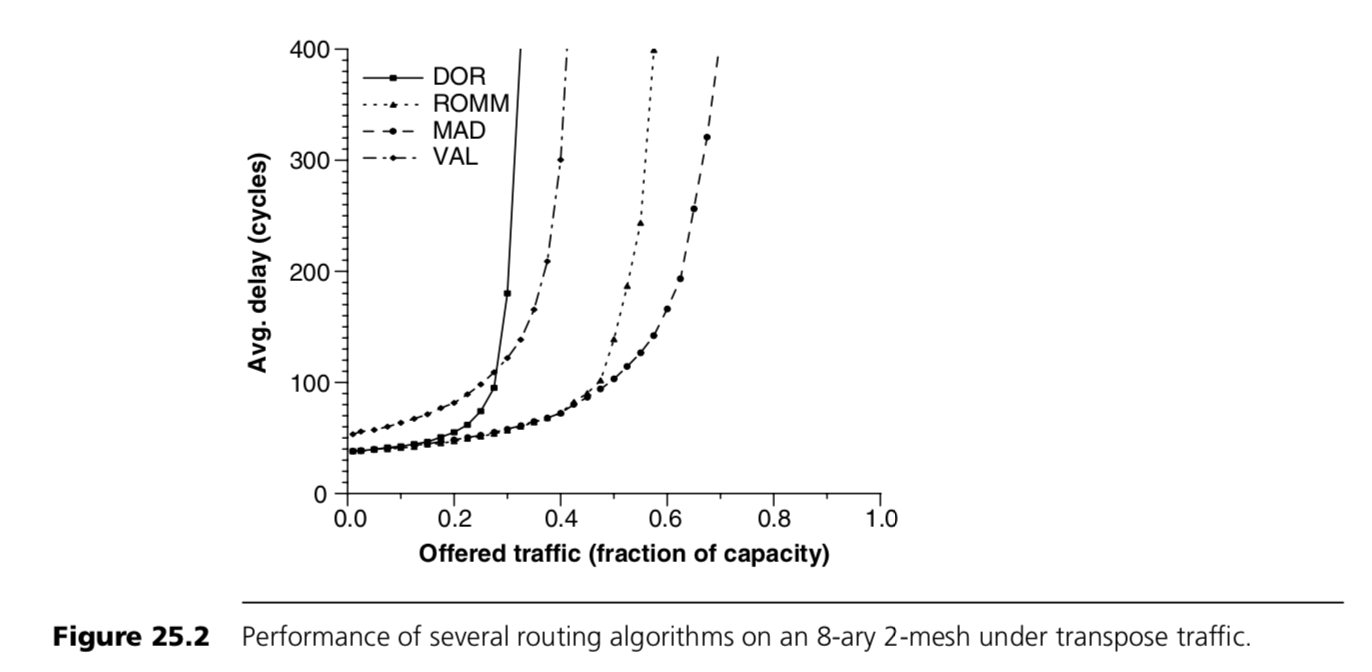}}
\end{figure}

\begin{figure}[tb]
\centerline{\includegraphics[width=\linewidth]{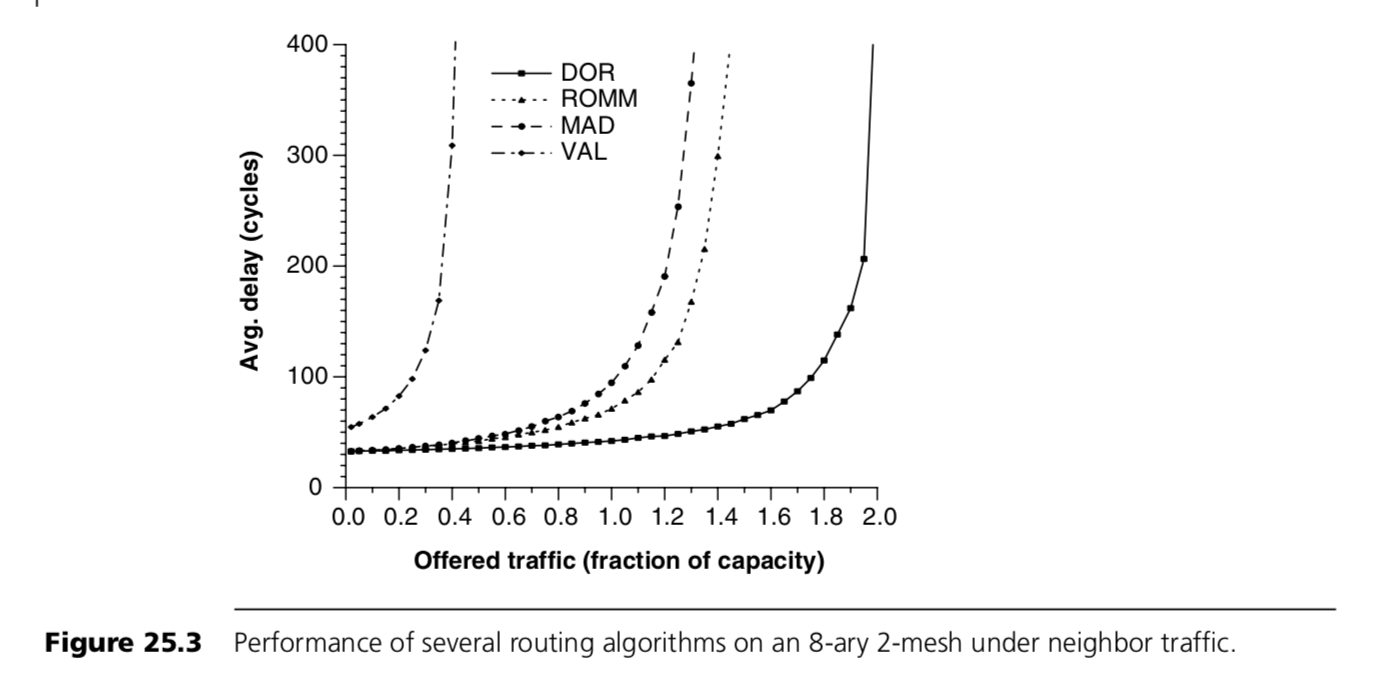}}
\end{figure}

\bibliography{sample}

\begin{thebibliography}{}

\bibitem[Davidson et~al., 2018]{Davidson_Celerity_IEEE_Micro_2018}
Davidson, S., Xie, S., Torng, C., Al-Hawaj, K., Rovinski, A., Ajayi, T., Vega,
  L., Zhao, C., Zhao, R., Dai, S., Amarnath, A., Veluri, B., Gao, P., Rao, A.,
  Liu, G., Gupta, R.~K., Zhang, Z., Dreslinski, R., Batten, C., and Taylor,
  M.~B. (2018).
\newblock {The Celerity Open-Source 511-core RISC-V Tiered Accelerator Fabric}.
\newblock {\em Micro, IEEE}.

\bibitem[Taylor, 2007]{taylor2007tiled}
Taylor, M.~B. (2007).
\newblock {\em Tiled microprocessors}.
\newblock PhD thesis, Massachusetts Institute of Technology, Department of
  Electrical Engineering and Computer Science.

\bibitem[Taylor, 2018]{taylor_basejump}
Taylor, M.~B. (2018).
\newblock Basejump stl: systemverilog needs a standard template library for
  hardware design.
\newblock In {\em Design Automation Conference}.

\end{thebibliography}

\end{document}